\documentclass[aps, prx, twocolumn, floatfix, superscriptaddress, reprint, 10pt]{revtex4-1}  
\usepackage{bm}
\usepackage{graphicx}
\usepackage[usenames]{color}
\usepackage{microtype}
\usepackage{amsmath}
\usepackage{amssymb}
\usepackage{xspace}
\usepackage{footnote}
\usepackage{relsize}
\usepackage{hhline}

\usepackage{physics}
\usepackage{mhchem}
\usepackage{comment}
\usepackage[normalem]{ulem}

\newcommand{\br}{\mathbf{r}}

\newcommand{\Rhat}{{\hat{R}}}
\newcommand{\ihat}{{\hat{i}}}

\newcommand{\bx}{\mathbf{x}}

\newcommand{\bP}{\mathbf{P}}
\newcommand{\bmu}{\boldsymbol{\mu}}

\makeatletter
\let\oldket\ket 
\def\ket{\@ifstar{\oldket}{\oldket*}}
\let\oldbra\bra
\def\bra{\@ifstar{\oldbra}{\oldbra*}}
\let\oldev\ev
\def\ev{\@ifstar{\oldev}{\oldev*}}
\makeatother
\newcommand{\rrho}[2]{\overline{\rho_{#2}^{\otimes #1}}}
\newcommand{\rrhoi}[1]{\rrho{#1}{i}}

\newcommand{\rrhovi}{\overline{\rho_i \otimes V_i}}

\usepackage{xparse}
\NewDocumentCommand{\fket}{s m}{ \IfBooleanTF{#1}{\oldket{#2}}{\oldket*{#2}}}
\usepackage{xparse}
\NewDocumentCommand{\fbra}{s m}{ \IfBooleanTF{#1}{\oldbra{#2}}{\oldbra*{#2}}}

\newcommand{\D}[2][]{\ensuremath{\mathop{}\!\mathrm{d}^{#1}{#2}\,}}

\newcommand{\Othree}{{O(3)}}
\newcommand{\SOthree}{{SO(3)}}

\newcommand{\rcut}{r_\text{c}}
\newcommand{\lmax}{l_\text{max}}

\newcommand{\elem}{a}

\begin{document}

\title{
Multi-scale approach for the prediction of atomic scale properties
}

\author{Andrea Grisafi}
\thanks{These authors contributed equally to this work}
\affiliation{Laboratory of Computational Science and Modeling, IMX, \'Ecole Polytechnique F\'ed\'erale de Lausanne, 1015 Lausanne, Switzerland}

\author{Jigyasa Nigam}
\thanks{These authors contributed equally to this work}
\affiliation{Laboratory of Computational Science and Modeling, IMX, \'Ecole Polytechnique F\'ed\'erale de Lausanne, 1015 Lausanne, Switzerland}
\affiliation{Indian Institute of Space Science and Technology, Thiruvananthapuram 695547, India}
\affiliation{National Centre for Computational Design and Discovery of Novel Materials (MARVEL), {\'E}cole Polytechnique F{\'e}d{\'e}rale de Lausanne, 1015 Lausanne, Switzerland}

\author{Michele Ceriotti}
\email{michele.ceriotti@epfl.ch}
\affiliation{Laboratory of Computational Science and Modeling, IMX, \'Ecole Polytechnique F\'ed\'erale de Lausanne, 1015 Lausanne, Switzerland}
\affiliation{National Centre for Computational Design and Discovery of Novel Materials (MARVEL), {\'E}cole Polytechnique F{\'e}d{\'e}rale de Lausanne, 1015 Lausanne, Switzerland}

\begin{abstract}
Electronic nearsightedness is one of the fundamental principles governing the behavior of condensed matter and supporting  its description in terms of local entities such as chemical bonds. 
Locality also underlies the tremendous success of machine-learning schemes that predict quantum mechanical observables -- such as the cohesive energy, the electron density, or a variety of response properties -- as a sum of atom-centred contributions, based on a short-range representation of atomic environments.
One of the main shortcomings of these approaches is their inability to capture physical effects, ranging from electrostatic interactions to quantum delocalization, which have a long-range nature. 
Here we show how to build a multi-scale scheme that combines in the same framework local and non-local information, overcoming such limitations. We show that the simplest version of such features can be put in formal correspondence with a multipole expansion of permanent electrostatics. The data-driven nature of the model construction, however, makes this simple form suitable to tackle also different types of delocalized and collective effects. We present several examples that range from molecular physics, to surface science and biophysics, demonstrating the ability of this multi-scale approach to model interactions driven by electrostatics, polarization and dispersion, as well as the cooperative behavior of dielectric response functions.
\end{abstract}

\maketitle

\section{Introduction}

The broad success of machine-learning approaches, used to predict atomic-scale properties bypassing the computational cost of first-principles calculations~\cite{behl-parr07prl,bart+10prl,khal+10prb,eshe+12prl,soss+12prb,mora+16pnas,deri-csan17prb,chmi+18nc,schu+18jcp,welb+18jctc,fabe+18jcp,zhan+18prl,drag+18prm,chen+19pnas}, can be largely traced to the use of structural descriptors that are defined through localized atomic environments~\cite{behl11pccp,bart+13prb,will+19jcp}. 
The assumption of locality is supported by the principle of nearsightedness of electronic matter first introduced by Walter Kohn~\cite{prod-kohn05pnas}, which implies that far-field perturbations to the local properties of the system are usually screened, and exponentially-decaying. 
Locality has been long exploited to develop linear-scaling electronic-structure schemes~\cite{yang91prl,gall-parr92prl,kohn96prl,pals-mano98prb,goed99rmp}, and in the context of machine-learning methods it allows constructing models that are highly transferable, and applicable to diverse problems as well as to complex, heterogeneous datasets~\cite{bart+17sa}.

Structural descriptors that are built using only local information cannot, however, describe long-range interactions and non-local phenomena. 
In many contexts, particularly when describing homogeneous, bulk systems~\cite{mora+16pnas}, long-range tails can be incorporated in an effective way, or approximated by increasing the range of the local environments~\cite{nata+15pccp}. On a fundamental level, however, the use of nearsighted representations undermines the reliability of machine-learning approaches whenever strong electrostatic and polarization effects guide the macroscopic behavior of the system. 
This is for instance the case when considering the electrostatic screening properties of water and electrolyte solutions~\cite{zhan-gall14jcp,Smith2016,chen+16sa,gaid-gall17jpcl,bell+18jpcl,Coupette2018}, the collective dispersion interactions that stabilize molecular crystals and biomolecules~\cite{reil-tkat14prl,ambr+16science}, or the surface charge polarization of a metal electrode in response to an external electric field~\cite{Siepmann1995,Merlet2013,Dufils2019,Scalfi2020,Elliott2020}.
Several examples have been presented that demonstrate the shortcomings of local ML models in the presence of long-range physical effects~\cite{gris+18prl,wilk+19pnas,veit+20jcp}.

\begin{figure*}[tbhp]
    \centering
\includegraphics[width=1.0\linewidth]{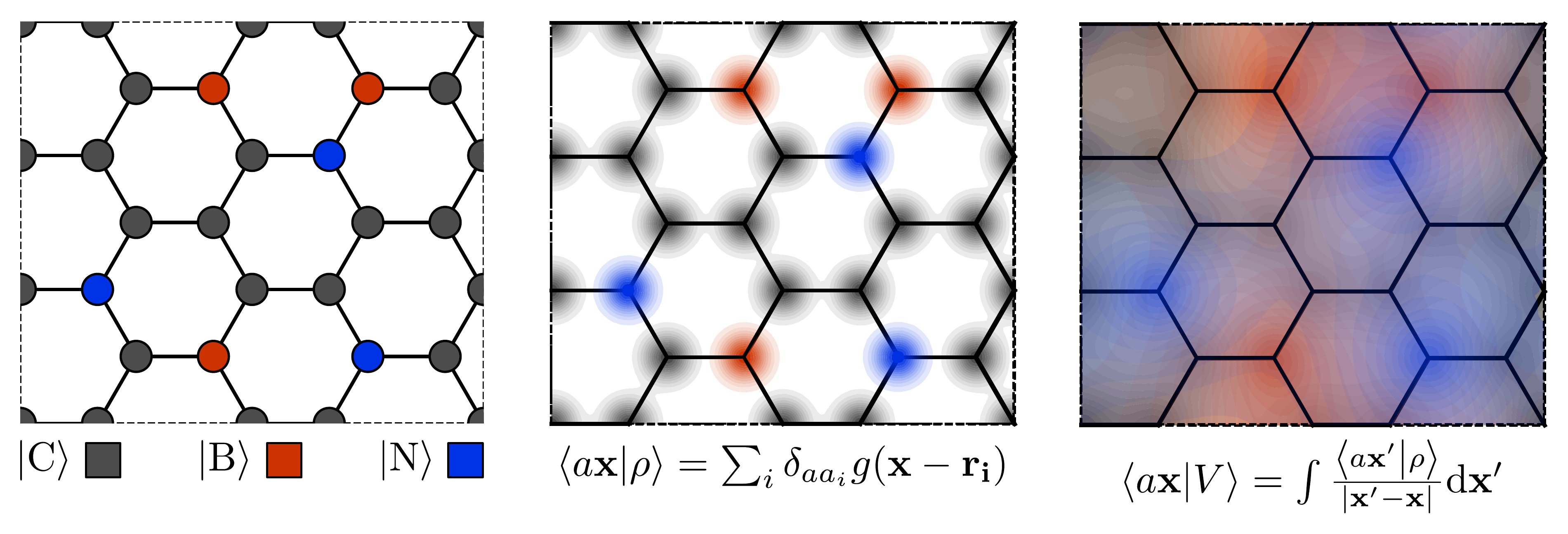}
\caption{Schematic representation of the construction of atomic-field representations. Left: Cartesian coordinate of the atoms for a hypothetical ``doped graphene'' system; middle: atom-density field, divided in three elemental channels, color-coded; right: atom-density potential, color-coded.}
    \label{fig:fields}
\end{figure*}

Global representations exist that incorporate information on the entire system~\cite{rupp+12prl,Huo2017arxiv,Hirn2017}, but usually they reduce the transferability of the resulting model. In the context of modelling electronic potential energy surfaces, several strategies have been proposed to incorporate explicitly the physical effects that underlie long-range interactions. Baselining the model with a cheaper electronic-structure method that incorporates electrostatic contributions~\cite{rama+15jctc,bart+10prl,welb+18jctc,Deng2019,ross+20jctc}, or using free-energy perturbation to promote a short-range ML potential to full quantum chemical accuracy~\cite{chen+19pnas} are very effective, pragmatic approaches to circumvent the problem. 
Alternatively, one can directly machine-learn the atomic partial charges and multipoles that enter the definition of the electrostatic energy~\cite{hand-pope09jctc,artr+11prb,bere+15jctc,Bereau2017,Bleiziffer2018,Nebgen2018,Yao2018,Zhang2019arxiv}, the atomic polarizability that underlies dispersion interactions~\cite{bere+18jcp}, or atomic electronegativities that are then used to determine the  partial charges of the system by minimizing its electrostatic energy~\cite{ghas+15prb,Faraji2017}. 
The major shortcoming of these methods is that, on one side, they are highly system dependent and, on the other, they are limited to the prediction of energy-related properties, and to the specific physical interaction that they are designed to model.
Some of the present Authors have recently proposed an alternative approach to incorporate non-local interactions into an atom-centred ML framework. Non-local information of the system is folded within local atomic environments thanks to the definition of smooth Coulomb-like potentials that are subsequently symmetrized according to the nature of the target property~\cite{gris-ceri19jcp}. 
The resulting long-distance equivariant (LODE) representation is endowed with a long-range character while still being defined from the information sampled in a finite local neighbourhood of the atoms. 

In this work, density and potential based descriptors are combined within a unified multi-scale representation. The resulting model can be formally related to an environment-dependent multipolar expansion of the electrostatic energy, but has sufficient flexibility to yield accurate predictions for a number of different kinds of interactions, and regression targets. 
We first consider, as an example, a dataset of organic dimers, partitioned into pairs that are representative of the possible interactions between charged, polar and apolar groups, demonstrating that the multi-scale LODE features can be used to describe permanent electrostatics, polarization and dispersion interactions with an accuracy that is only limited by the number of training points. 
We then show how our model is able to capture the mutual polarization between a water molecule and a metal slab of lithium. Finally, we reproduce the dipole polarizability of a dataset of poly-aminoacids, extrapolating the electric response of the system at increasing chain lengths.

\section{Multi-scale equivariant representations}

Following the notation introduced in Ref.~\citenum{will+19jcp},  
let us start by defining a density field associated with a structure $A$ as the superposition of decorated atom-centred Gaussians  
\begin{equation}
\label{eq:density}
\bra{\elem \bx}\ket{A; \rho}=
\sum_{i\in A} \delta_{\elem\elem_i} g(\bx - \br_i)\,.
\end{equation}
In this expression, $\br_i$ indicates the position of atoms of $A$, and $\elem_i$ labels their elemental nature. From these smooth atomic densities, a Coulomb-like potential can be formally defined as a result of the following integral operation:
\begin{equation}
\label{eq:potential}
\bra{\elem \bx}\ket{A; V}= \int d\bx' \frac{\bra{\elem \bx'}\ket{A; \rho}}{|\bx-\bx'|}\, .
\end{equation}
One could build a general family of fields using a different integral transformation of the density, but here we focus on this $1/|\bx-\bx'|$ form, which is well-suited to describe long-range interactions.
The two primitive representations $\ket{\rho}$ and $\ket{V}$ can be individually symmetrized over the continuous translation group~\cite{will+19jcp}. Imposing translational invariance on Eq.~\eqref{eq:density} has the ultimate effect of centring the representation on the atoms $i$ of the system, so that we can conveniently refer to the set of atom-centred densities\footnote{Strictly speaking, $g$ in Eq.~\eqref{eq:density-centred} has twice the variance as that in~\eqref{eq:density}, but we re-define the density function accordingly. }
\begin{equation}
\label{eq:density-centred}
    \bra{\elem \bx}\ket{A; \rho_i} = \sum_{j\in A}\delta_{\elem\elem_j}  g(\bx - \br_{ji})\,,
\end{equation}
where $\br_{ji}=\br_j-\br_i$ is the vector separating atoms $i$~and~$j$.
As already shown in Ref.~\citenum{gris-ceri19jcp}, given that the Coulomb operator is translationally invariant, one can obtain an analogous result symmetrizing the tensor product $\ket{\rho}\otimes\ket{V}$, yielding a set of atom-centred potentials
\begin{equation}
\label{eq:potential-centred}
\bra{\elem \bx}\ket{A; V_i}= \int d\bx' \frac{ \sum_{j\in A} \delta_{\elem\elem_j} g(\bx' - \br_{ji})}{|\bx-\bx'|}\, .
\end{equation}

Either of Eqs.~\ref{eq:density-centred} or~\ref{eq:potential-centred} contains information on the entire structure. Usually, however, the atom-centred density $\ket{\rho_i}$ is evaluated including only atoms within spherical environments of a given cutoff radius~$\rcut$. This truncation is not only a matter of practical convenience: fundamental physical principles~\cite{prod-kohn05pnas} indicate that molecular and materials properties are largely determined by local correlations, and increasing indefinitely $\rcut$ has been shown to \emph{reduce} the accuracy of the model~\cite{bart+17sa,will+18pccp}, because, in the absence of enormous amounts of uncorrelated training structures, the increase in model flexibility leads to overfitting. 
The fundamental intuition in the construction of the atom-density potential $\ket{V_i}$ is that, even if one evaluates it in a spherical  neighbourhood of the central atom $i$, thereby avoiding an uncontrollable increase in the complexity of the model, it incorporates contributions from atoms that are very far away. 
The nature of $\ket{V_i}$ can  be better understood by separating the near-field from the far-field potential in the definition of Eq.~\eqref{eq:potential-centred}, that is,
\begin{equation}\label{eq:split}
\begin{split}
\bra{\elem \bx}\ket{V_i} &=\bra{\elem \bx}\ket{V_i^<} + \bra{\elem \bx}\ket{V_i^>} = \\& = \int d\bx' \frac{\bra{\elem \bx'}\ket{\rho^<_i}}{|\bx-\bx'|} + \int d\bx' \frac{\bra{\elem \bx'}\ket{\rho^>_i}}{|\bx-\bx'|}
\end{split}
\end{equation} 
where $\rho^<_i$ and $\rho^>_i$ are the atomic densities located inside and outside the $i$-th spherical environment. We omit the structure label $A$ for convenience, as we will do often in what follows. The near-field term contributes information that is analogous to that included in $\ket{\rho_i}$. The far-field contribution instead determines the effect of the density beyond $\rcut$, and the choice of the integral operator affects the asymptotic form of this effect, with $1/|\bx-\bx'|$ implying a Coulomb-like behavior.

Tensor products of the atom-centred density Eqs.~\eqref{eq:density-centred} and potential ~\eqref{eq:potential-centred} could be separately symmetrized over rotations and inversion, yielding respectively structural descriptors of short-range interatomic correlations, equivalent to SOAP-like representations~\cite{bart+13prb}, or long-distance equivariants (LODE) features~\cite{gris-ceri19jcp}.
Here we introduce a more explicitly multi-scale family of representations, that couples $\ket{\rho_i}$ and $\ket{V_i}$ terms. Formally, one can obtain a symmetry-adapted ket that transforms like the irreducible representations of the $\Othree$ group by computing the  Haar integral over the rotation and inversion operators $\Rhat$, $\ihat$:
\begin{multline}
\label{eq:equivariant}
\ket{\ev{\rho_i^{\otimes \nu}\otimes V_i^{\otimes \nu'}\otimes \sigma \otimes \lambda\mu}_\Othree} =  \sum_{k=0,1}\int_\SOthree\!\!\!\!\!\!\!\!\D{\Rhat}\ihat^k\ket{\sigma}  \otimes \\
\ihat^k\Rhat\ket{\lambda\mu} \otimes\underbrace{\ihat^k\Rhat\ket{\rho_i} \otimes\ldots 
\ihat^k\Rhat\ket{\rho_i} }_{\nu\ \text{times}}\otimes
\underbrace{\ihat^k\Rhat\ket{V_i} \otimes\ldots 
\ihat^k\Rhat\ket{V_i} }_{\nu'\ \text{times}}\,,
\end{multline}
which we indicate in what follows using the shorthand notation $\ket{\overline{\rho_i^{\otimes \nu}\otimes V_i^{\otimes \nu'};\sigma;\lambda\mu}}$, omitting the $\sigma;\lambda\mu$ indices when considering invariant features ($\sigma=1$, $\lambda=0$).
Within this construction, the ket $\ket{\lambda\mu}$ has the role of making the resulting features transform as a $Y_\lambda^\mu$ spherical harmonic~\cite{gris+18prl,gris+19book}, while $\ket{\sigma}$ indicates the parity of the features under inversion~\footnote{In particular, we consider $\sigma=1$ if the learning target behaves as a polar tensor and $\sigma=-1$ if it mimics a pseudotensor under inversion symmetry.}. 

In practical implementations, the abstract ket~\eqref{eq:equivariant} can be computed by first expanding the atom-centred features~\eqref{eq:density-centred}-\eqref{eq:potential-centred} onto a discrete basis, and then evaluating the symmetrized $\nu$-point correlation of the fields. A particularly clean, efficient, recursive formulation can be derived exploiting the fact that the equivariant features behave as angular momenta, and can then be combined using Clebsch-Gordan coefficients to build higher-order correlations~\cite{nigam2020recursive}.
In analytical derivations we use a partially-discretized basis, in which the radial contribution is kept as a continuous index, corresponding to 
\begin{equation}\label{basis}
\bra{\elem rlm} = \int\D{\bx}  \delta(r-x) Y_l^m(\hat{\bx}) \bra{\elem\bx},
\end{equation}
while in the actual implementation we use a basis of Gaussian type orbitals to also discretize the radial component.

The nature of the representation, however, does not depend on such details. The basis-set independence is most clearly seen by considering the use of the equivariants in the context of a linear regression model.
The value of a tensorial property $\mathbf{T}$ for a structure, decomposed in its irreducible spherical components (ISC)~\cite{Stone} and in atom-centred contributions, can be formulated as 
\begin{equation}
\label{eq:tensors}
T^{\sigma\lambda}_{\mu}(A,i) \approx\int\D{X} \bra{T;\sigma\lambda}\ket{X} \bra{X}
\ket{A; \overline{\rho_i^{\otimes \nu}\otimes V_i^{\otimes \nu'}; \sigma; \lambda\mu}},
\end{equation}
where $X$ indicates any complete basis that provides a concrete representation of the ket, and $\bra{X}\ket{T;\sigma\lambda}$ is the set of regression weights. 
One sees that (1) the regression model is invariant to a unitary transformation of the basis; (2) the equivariant transformation of the target property is associated with the $\lambda\mu$ indices of the ket, while the weights are invariant under symmetry operations.
Linear models are especially useful to reveal the physical meaning of a representation: they allow to demonstrate the relation between short-range density correlations ($\nu'=0$) and the body-order expansion of interatomic potentials~\cite{glie+18prb,will+19jcp,drau19prb,jinn+20jcp}, and the relation between the first-order LODE ($\nu=0, \nu'=1$) and fixed point-charge electrostatics.
In the next section, we use this idea to show how the simplest multi-scale LODE ($\nu=1, \nu'=1$) can be put in formal correspondence with the physics of multipole electrostatics~\cite{stone1997}.

\section{Analytical results for long-range interactions}

Consider a linear model to  predict the ground-state electronic energy $U$ of a system $A$. This corresponds to taking the scalar ($\lambda=0$) and polar ($\sigma=1$) limits within the prediction formula of Eq.~\eqref{eq:tensors}:
\begin{equation}\label{eq:epred}
U(A) =\sum_{i=1}^N U_i(A) =\sum_{i=1}^N \int\D{X} \bra{U}\ket{X} \bra{X}\ket{A; \rrhovi}.
\end{equation}
We aim to prove that in the  LODE(1,1) case, where the density and potential representations are both introduced to first order, this functional form can be used to model rigorously a multipolar expansion of the long-range contributions to  $U$. 

To see this, let us start by representing the energy prediction in terms of the partially-discretized basis of Eq.~\eqref{basis}. Upon symmetrization of the tensor product between $\rho$ and $V$, and going to the coupled angular momentum basis~\cite{nigam2020recursive}, one obtains a set of invariants that can be expressed using the basis $\bra{X}\equiv\bra{\elem_1 r_1;\elem_2 r_2; l}$
\begin{multline}
\bra{\elem_1 r_1; \elem_2 r_2; l}\ket{\rrhovi} = \frac{1}{\sqrt{2l+1}} \sum_{|m|\le l} \\ \bra{\elem_1 r_1 l m}\ket{\rho^<_i} \bra{\elem_2 r_2 l m}\ket{V_i}^\star,\label{eq:rhov-rr}
\end{multline}
where $\bra{\elem_1 r_1 l m}\ket{\rho^<_i}$ and $\bra{\elem_2 r_2 l m}\ket{V_i}$ indicate the spherical harmonics projections of the local density and the total potential fields respectively. 
Focusing in particular on the contribution to the energy originating from the long-range part of the potential field around atom $i$, $\ket{V_i^>}$ (cf. Eq.~\eqref{eq:split}), we can write explicitly Eq.~\eqref{eq:epred} as follows:
\begin{multline}\label{eq:energy-basis}
U^>_i = \sum_{l=0}^{\lmax} \sum_{\elem_1\elem_2}\int_0^{\rcut} \D{r_1} \int_0^{\rcut} \D{r_2} \bra{U}\ket{\elem_1 r_1; \elem_2 r_2; l} \\ \times  \sum_{|m|\le l} \bra{\elem_1 r_1 l m}\ket{\rho^<_i} \bra{\elem_2 r_2 l m}\ket{V_i^>}^\star\, .
\end{multline}
where $\bra{U}\ket{\elem_1 r_1;\elem_2 r_2; l}$ indicates the regression weights for the \emph{total} potential energy, that also incorporate the  ${1}/{\sqrt{2l+1}}$ factor in Eq.~\eqref{eq:rhov-rr}. 
We are now interested in representing the spherical harmonic components of the potential in terms of the far-field contribution $V^>_i$ of Eq.~\eqref{eq:split}. Using the Laplace expansion of the Coulomb operator, we can rewrite $\ket{V_i^>}$ as :
\begin{multline}\label{eq:sph-potential}
\bra{\elem_2 r_2 l m}\ket{V^>_i}^\star= \frac{4 \pi}{2l+1} \int \D{\bx} \, \bra{\rho^>_i}\ket{\elem_2\bx}  \frac{r_2^l}{x^{l+1}} \bra{\hat{\bx}}\ket{lm}\\
=\frac{4 \pi}{2l+1}\int \D{r} \bra{\rho^>_i}\ket{\elem_2 r l m} \frac{r_2^l}{r^{l+1}}   .
\end{multline}
Plugging this into Eq.~\eqref{eq:energy-basis}, one sees that the contribution to the energy coming from the far-field can be written as 
\begin{equation}
\label{eq:ulr-multipoles}
U^>_i = \sum_{\elem_1\elem_2}\sum_{lm} \int_{\rcut}^{\infty} \!\!\!\D{r} \frac{1}{r^{l+1}} \bra{\rho^>_i}\ket{\elem_2 r l m}
\bra{\elem_1 \elem_2; l m}\ket{M_i^<(U)}\,,
\end{equation}
where we introduced a set of regression-driven spherical multipoles 
\begin{multline}\label{eq:multipoles}
\bra{\elem_1 \elem_2; l m}\ket{M_i^<(U)} =\\ \frac{4 \pi}{2l+1} \int_0^{\rcut} dr_2\, r_2^l  \int_0^{\rcut} dr_1\,  \bra{U}\ket{\elem_1 r_1;\elem_2 r_2; l}\bra{\elem_1 r_1 l m}\ket{\rho^<_i} .
\end{multline}
Eq.~\eqref{eq:ulr-multipoles} shares a striking resemblance with the expression for the interaction of a far-field charge density with the electrostatic potential generated by the near-field charge distribution~\cite{Griffiths2013}. As we shall see in what follows that this formal equivalence underpins the ability of $\ket{\rrhovi}$ to model accurately several kinds of interactions. 
Crucially, however, $\rho_i$ and $V_i$ do not represent physical quantities, but are just a representation of the spatial arrangement of atoms. Atoms in the far-field respond in a way that depends only on their chemical nature, but the local multipoles are modulated in a highly flexible, non-trivial fashion by the distribution of atoms in the local environment.
The form of Eq.~\eqref{eq:multipoles} also hints at how changing the representation would affect this derivation. Increasing the density order $\nu$ would allow for a more flexible, higher-body-order dependence of the local multipoles on the distribution of atoms in the vicinity of atom~$i$, while increasing $\nu'$ would bring a more complicated dependency on the distribution of atoms in the far-field, leading to a linear regression limit that does not match formally the electrostatic multipole expansion. Changing the asymptotic form of the potential in Eq.~\ref{eq:potential} could be used to incorporate explicitly dispersion-like, $1/r^6$ features. 
We want to stress that even in this form the model is not limited to describing the physics of permanent electrostatics. In fact, the coupling between the inner and outer atomic species ($\elem_1$ and $\elem_2$) carried by the definition of the regression weights makes it possible for the local multipoles to respond to species of the far-field distribution. 
We test the limits of this data-driven approach in Section~\ref{sec:results}. 

\begin{figure}[tbhp]
    \centering
    \includegraphics[width=0.9\linewidth]{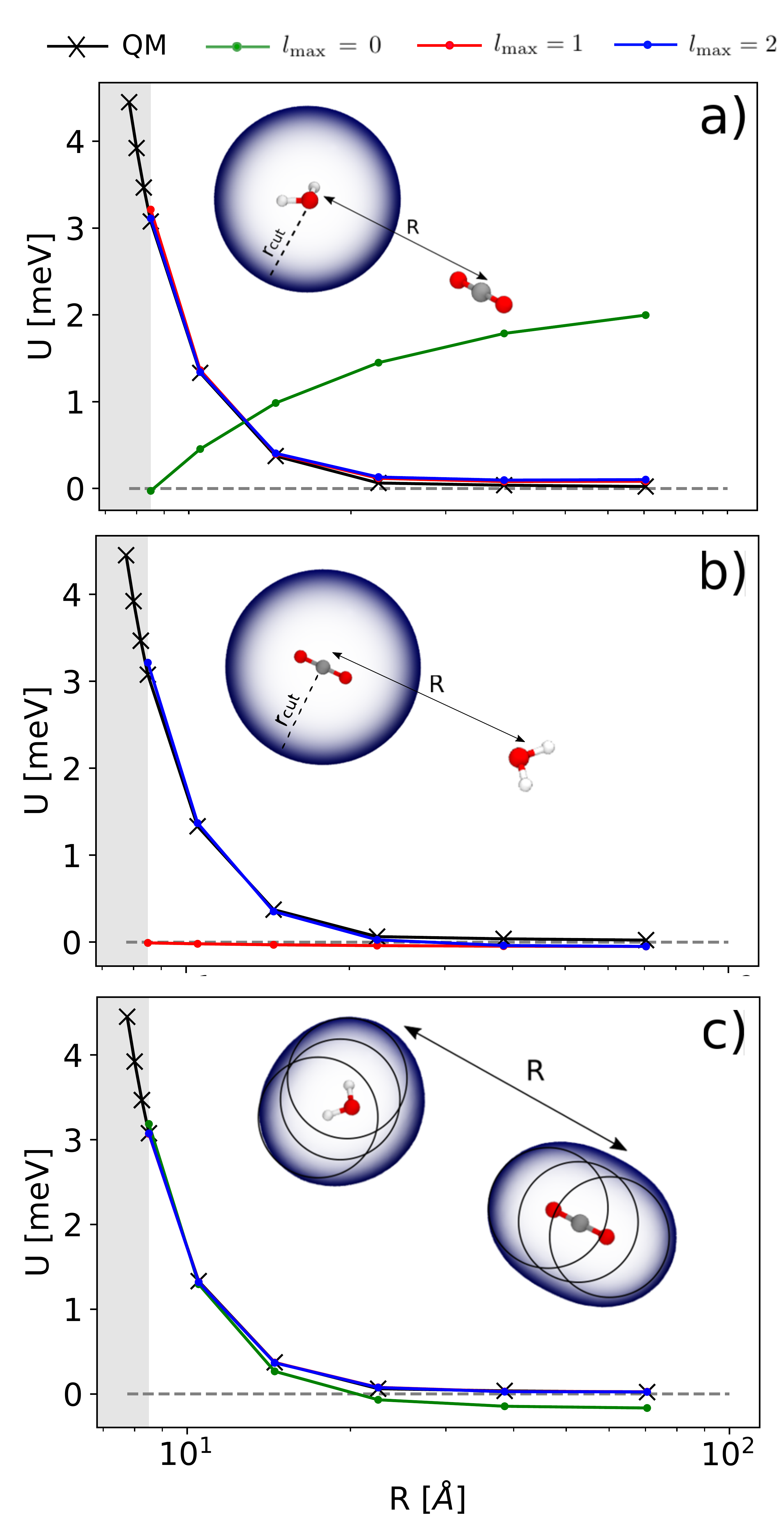}
    \caption{Extrapolated interaction profiles for a given configuration of H$_2$O and CO$_2$ at different angular cutoff values $\lmax$. Top, middle and bottom panels show the results of the asymptotic extrapolation when centring the representation on the oxygen atom of H$_2$O, the carbon atom of CO$_2$ and all the atoms of the system respectively.}
    \label{fig:multipoles}
\end{figure}

Before doing so, we want however to demonstrate numerically the mapping between these multi-scale LODE features and multipole models. 
According to our construction, the cutoff value $\lmax$ chosen to define the angular resolution of the representation determines the number of multipoles that are included within the expansion of Eq.~\eqref{eq:sph-potential}. For example, when truncating at $\lmax=0$ and taking the limits of vanishing Gaussian width $\sigma\to0$ and radial cutoff $\rcut\to0$,  Eq.~\eqref{eq:sph-potential} takes the form of a simple Coulomb interaction between point-changes~\cite{rupp+12prl} (the derivation is reported in the Supplementary Material). Including multipoles for $l>0$ makes it possible to represent the anisotropy of the electrostatic interaction.
We analyze quantitatively the behavior of a linear model based on $\ket{\rrhovi}$ by observing its performance in representing the far-field interactions between an \ce{H2O} and a \ce{CO2} molecule -- since the leading term of the interactions between the two molecules is driven by permanent electrostatics. 
We build a dataset considering 33 non-degenerate reciprocal orientations between the two molecules, and learn the interaction over a range of distances between the centres of mass from 6.5 to 9{\AA}. We then extrapolate the predicted interaction profile in the asymptotic regime of~$R>9${\AA}. 
In Fig~\ref{fig:multipoles} we report the results of the extrapolation for a given reciprocal orientation at increasing angular cutoffs $\lmax$. We also compare different choices for the possible atomic centres that contribute to the energy prediction: in panel (a) we express the energy in terms of a single environment centred on the oxygen atom of the \ce{H2O} molecule; in panel (b) we use a single environment centred on the carbon atom of \ce{CO2}; in (c) we use multiple environments centred on each atom. This exercise probes the possibility of choosing between a model for the electrostatic energy that is based on the definition of \textit{molecular} rather than \textit{atomic} multipoles~\cite{bere+15jctc,Bereau2017}.

As one would expect from a classical interpretation of the long-range energy, the binding profile for the selected test configuration is ultimately driven by the interaction between the dipole moment of the water molecule and the quadrupole moment of CO$_2$. This is reflected in the sharp transition of the prediction accuracy when crossing a critical angular cutoff $\lmax$. When centring the local environment on the water molecule (Fig.~\ref{fig:multipoles}-a)), for instance, truncating the expansion at $\lmax=1$ is enough to reproduce the interaction between the dipolar potential of water and the \ce{CO2} molecule. Conversely, when centring the representation on carbon dioxide (Fig.~\ref{fig:multipoles}-b)), the \ce{H2O} density in the far-field has to interact with a \ce{CO2} potential that is quadrupolar in nature, which requires an angular cutoff of at least $\lmax=2$. 
When centring the representation on all the atoms of the system (Fig.~\ref{fig:multipoles}-c)), using an angular cutoff of $\lmax=0$ suffices to obtain qualitatively accurate interaction profiles. This is consistent with the success of the many parametrized force fields and machine-learning potentials that simply rely on representing the electrostatic energy of the system via a set of point-charges~\cite{artr+11prb,ghas+15prb}.
For this simple toy problem, increasing the expansion at $\lmax=1$ with an atomic multipole model achieves almost perfect predictions. 
Results of all of the 33 orientations we considered are reported in the SM.

\section{Results}
\label{sec:results}

The toy system we have discussed in the previous section reflects the  behavior of the multi-scale LODE representation. However, if $\ket{\rrhovi}$ was only capable of describing permanent electrostatics there would be little advantage over traditional physics-based models. 
In this Section we present three applications to substantially more complicated systems, to demonstrate that even in their simplest form, this family of features is suitable to address the complexity of challenging, real-life atomistic modelling problems.
We report errors in terms of the root mean square error (RMSE), or the percentage RMSE (RMSE\%), which is expressed as a percentage of the standard deviation of the target properties. 

\begin{figure*}
    \centering
    \includegraphics[width=18cm]{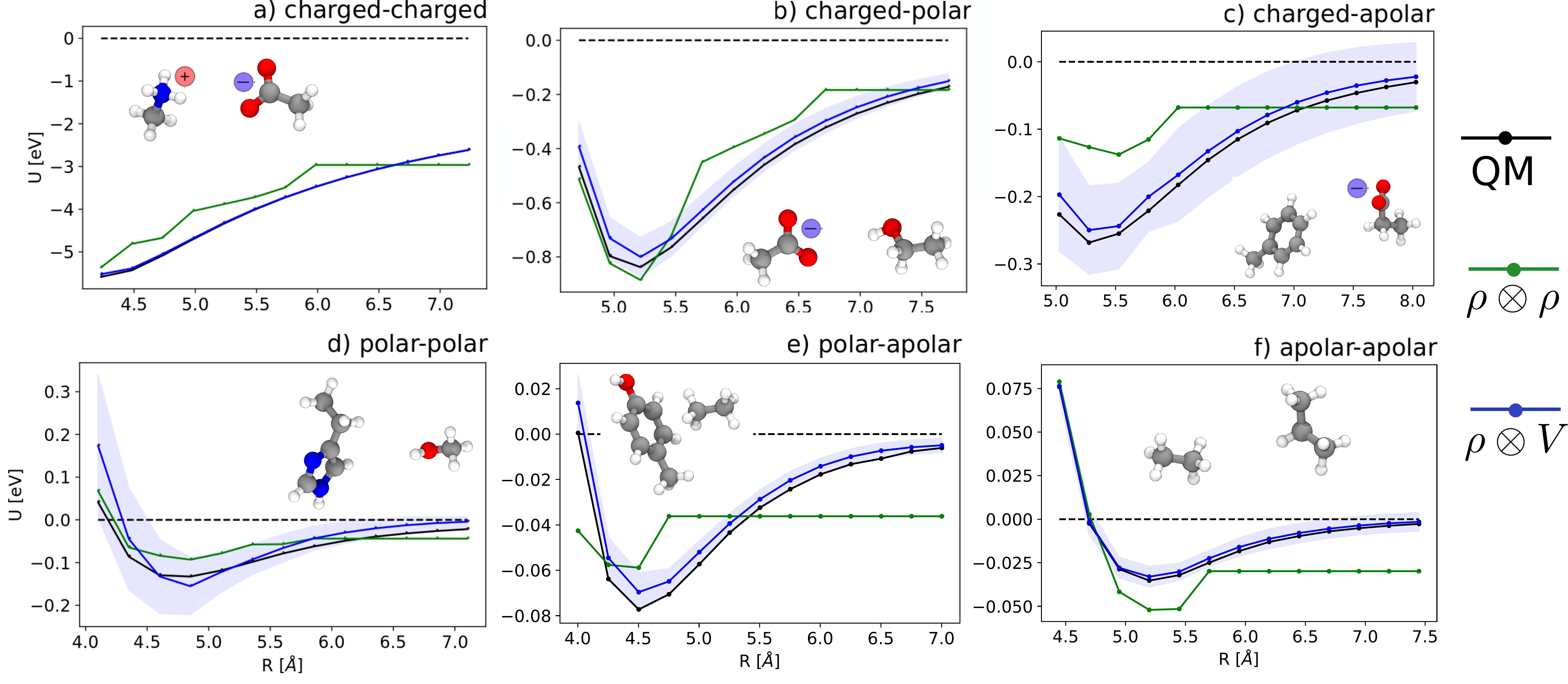}
    \caption{Median-error binding curves for six different classes of intermolecular interactions. (\textit{black lines}) quantum-mechanical calculations. (\textit{green lines}) predictions of a $\fket{\overline{\rho_i\otimes \rho_i}}$ model. (\textit{blue lines})  predictions of a $\fket{\rrhovi}$ model.}
    \label{fig:dimers}
\end{figure*}

\subsection{Binding energies of organic dimers}

We start by testing the ability of multi-scale LODE to describe different kinds of molecular interactions. To this end, we consider the interaction energy between 2291 pairs of organic molecules belonging to the BioFragment Database (BFDb)~\cite{Burns2017}. For each dimer configuration, binding curves are generated by considering 12 rigid displacements in steps of 0.25{\AA} along the direction that joins the geometric centres of the two molecules.  Then, unrelaxed binding energies are computed at the DFT/PBE0 level using the Tkatchenko-Scheffler self-consistent van der Waals method~\cite{Tkatchenko2009} as implemented in the FHI-aims package~\cite{Blum2009}. For each binding trajectory, we also include in the training set the dissociated limit of vanishing interaction energy, where the two monomers are infinitely far apart. The dataset so generated includes all the possible spectrum of interactions, spanning pure dispersion, induced polarization and permanent electrostatics. 
In order to better rationalize the learning capability of such a large variety of molecular interactions, we choose to partition the molecules in the dataset in three independent classes, namely, 1) molecules carrying a net charge, 2) neutral molecules that contain heteroatoms (\ce{N}, \ce{O}), and can therefore exhibit a substantial polarity 3) neutral molecules containing only \ce{C} and \ce{H}, that are considered apolar and interacting mostly through dispersive interactions.
Considering all the possible combinations of these kinds of molecules partitions the dimers into six classes, i.e., 184 charged-charged (CC), 267 charged-polar (CP), 210 charged-apolar (CA), 161 polar-polar (PP), 418 polar-apolar (PA) and 1051 apolar-apolar (AA) interactions. 
For each of the six classes, several, randomly selected binding curves are held out of the training set, to test the accuracy of our predictions. 
The remaining curves are used to fit one separate linear model for each class, using either local features (the SOAP power spectrum~\cite{bart+13prb}, $\ket{\rrhoi{2}}$) or multi-scale LODE($\nu=1,\nu'=1$) features. 
In order to also assess the reliability of our predictions, we use a calibrated committee estimator~\cite{musi+19jctc} for the model uncertainty, which allows us to determine error bars for the binding curves. 8 random subselections of 80\% of the total number of training configurations were are considered to construct the committee model. The internal validation set is then defined by selecting the training structures that are absent from at least 25\% of the committee members.  

\setlength{\tabcolsep}{0.5em}
\begin{table}[hbt]
\centering
\begin{tabular}{c|cccccc}
\hline\hline
& & & \multicolumn{3}{c}{RMSE/eV}\\
class & $n_\text{train}$  & STD/eV &  $\rho\otimes\rho$ & $\rho\otimes V$ & $V\otimes V$\\
\hline
CC & 100 & 1.86 & 0.72 & 0.049 & 0.058\\
CP &  200 & 0.379 & 0.25 & 0.074  & 0.092\\
CA &  150 & 0.083 & 0.056 & 0.041  & 0.034\\
PP &  100 & 0.131 & 0.10 & 0.062  & 0.125\\
PA &  350 & 0.046 & 0.032 & 0.013 & 0.021\\
AA &  950  & 0.063 & 0.026 & 0.004  & 0.006\\
\hline
\end{tabular}
\caption{Prediction performance expressed in terms of the RMSE over all the points of the binding curves, for the six classes of interactions and $\rho\otimes\rho$, $\rho\otimes V$ and $V\otimes V$ models. For each class we also indicate the number of training samples, and the characteristic energy scale, expressed in terms of the standard deviation of the energies in the test set. \label{tab:interactions}}
\end{table}

Figure~\ref{fig:dimers} shows characteristic interaction profiles for the six different classes of molecular pairs. The models use $\rcut=3$~{\AA} environments centred on each atom. The configurations we report are those that exhibit median integrated errors within the test set of each class. The root mean square errors associated with the predictions over the entire test sets of each class are listed in Table~\ref{tab:interactions}.
The results clearly show that while SOAP(2) is limited by the nearsightedness of the local environments, the LODE(1,1) multi-scale model is able to predict both the short and the long-range behaviour of the binding profiles on an equal footing. What is particularly remarkable is the fact that a simple, linear model can capture accurately different kinds of interactions, that occur on wildly different energy scales and asymptotic behavior: the typical binding energy of charged dimers is of the order of several eV, and has a $1/r$ tail, while the typical interaction energy of two apolar molecules is of the order of a few 10s of meV, and decays roughly as $1/r^6$. 
A LODE($\nu'=2$) model (i.e. based on $\ket{\overline{V_i^{\otimes 2}}}$ features) also allows to predict the binding curves beyond the 3\AA{} cutoff, but usually yields 50-100\%{} larger errors than those observed with $\ket{\rrhovi}$ -- not only for charged molecules, but also for dimers that are dominated by dispersion interactions. The multi-scale nature of LODE($\nu=1$,$\nu'=1$) yields a better balance of short and long-range descriptions, and is sufficiently flexible to be adapted to the description of systems that are not dominated by permanent electrostatics, even though interactions between charged fragments are considerably easier to learn, in comparison to the others.
We also observe that the uncertainty model works reliably, as the predicted curves always fall within the estimated error bar. Larger uncertainties are found for interaction classes that have few representative samples in the training set, such as those associated with polar-polar molecular pairs (Fig~\ref{fig:dimers}-d)). This observation reflects the fact that the model is limited by the training set size rather than by the nature of interaction involved, as confirmed by the lack of saturation~\cite{huan-vonl16jcp} observed in the learning curves (shown in the SM).

\subsection{Induced polarization on a metal surface}

The previous example proves that linear $\ket{\rrhovi}$ models  capture a wide class of molecular interactions, ranging from pure dispersion to permanent electrostatics. 
Beyond molecular systems, however, a large number of phenomena occur in solid state physics that are driven by long-range effects, and involve more subtle, self-consistent interactions between far-away atoms.
A particularly relevant example is represented by the induced macroscopic polarization that a metallic material undergoes in response to an external electric field, which underlies fundamentally and technologically important phenomena for surface science and nanostructures~\cite{bell+14jcp,litm+17jcp,maks+20ijqc}. Physics-based modelling of this kind of systems usually exploits the fact that, for a perfectly-conductive surface, the interaction is equivalent to that between the polar molecule and the mirror image, relative to the surface plane, of its charge distribution, with an additional inversion of polarity~\cite{finn+95jpcm}.
It would not appear at all obvious that our atom-centred framework, which does not include an explicit response of the far-field atom density to the local data-driven multipole, can capture the physics of a phenomenon associated with the polarization of electrons that are delocalized over the entire extension of the metallic solid.

\begin{figure}[tbp]
    \centering
    \includegraphics[width=8cm]{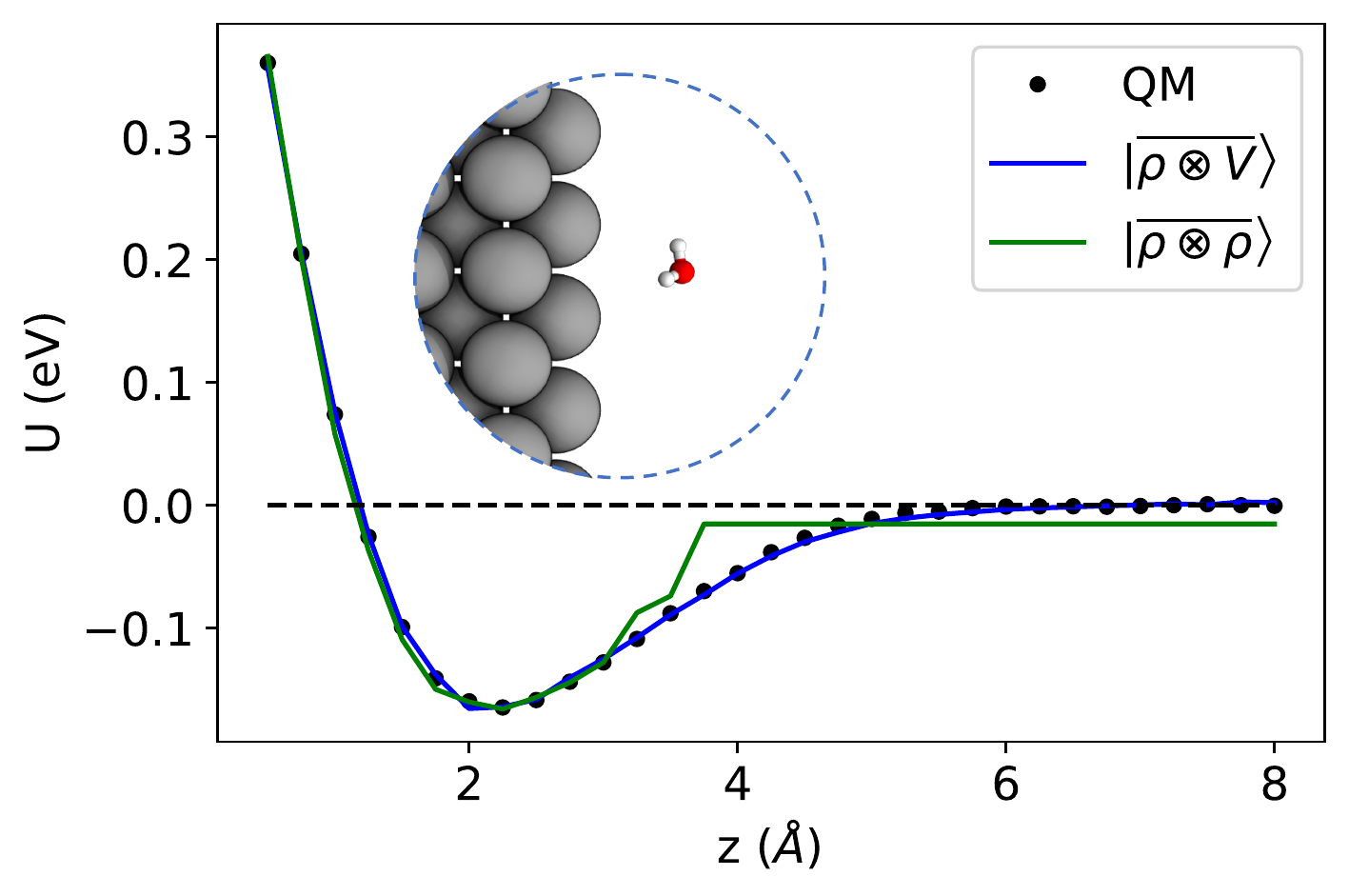}
    \caption{Predicted binding curve of a test water-lithium configuration. (\textit{black dots}) reference DFT calculations; (\textit{green line}) predictions of a $\fket{\overline{\rho_i \otimes \rho_i}}$ model; (\textit{blue line}) $\fket{\overline{\rho_i \otimes V_i}}$ model.}
    \label{fig:lithium-water}
\end{figure}

To benchmark the performance of multi-scale LODE in this challenging scenario we consider the interaction of a slab of \emph{bcc} lithium with a water molecule that is located at various distances from the (100)-surface. We start by selecting 81 water molecule configurations, differing in their internal geometry or in their spatial orientation relative to the surface. For each of these configurations, 31 rigid displacements are performed along the (100)-direction, spanning a range of distances between 0.5~\AA{} and 8~\AA{} from the lithium surface. 
Using this dataset we compute unrelaxed binding energies at the DFT/PBE level using the FHI-aims package~\cite{Blum2009}. We converge the slab size along the periodic $xy$-plane, minimizing the self-interaction between the periodic images of the water molecule, resulting in a $5\times5$ unit cell repetitions  and a $k$-points sampling of $4\times4\times1$  \AA$^{-1}$. We set the slab extension along the non-periodic $z$-direction so that the Fermi energy is converged within 10 meV, resulting in a total of 13 layers. To remove the spurious interactions along the $z$-axis, we set a large vacuum space of roughly 80~\AA{} in conjunction with a correction suitable to screen the dipolar potential~\cite{Neugebauer1992}. Following these prescriptions, we obtain attractive potential profiles for all molecular geometries and orientation, consistently with the interaction between the dipolar field of the water molecule and the induced metal polarization.

For this example, we construct $\ket{\rho_i}$ and $\ket{V_i}$ representations within spherical environments of $\rcut=4$~{\AA} with a Gaussian-density width of $\sigma=0.3$~{\AA}. The regression model is trained on 75 lithium-water binding curves while the remaining 6 are used for testing the accuracy of our predictions. Figure~\ref{fig:lithium-water} shows a comparison between a local $\ket{\rrhoi{2}}$ model and a multi-scale LODE $\ket{\rrhovi}$ model in learning the interaction energy of the metal slab and the water molecule for one representative test trajectory (all test trajectories are reported in the SM). %
We observe that the local SOAP description is able to capture the short-range interactions but becomes increasingly ineffective as the water molecule moves outside the atomic environment, leading to an overall error of about 19 RMSE\%{}. This is in sharp contrast to the performance of the $\ket{\rrhovi}$ representation, which can capture both the effects of electrostatic induction at a large distance and the Pauli-like repulsion at short range with the same level of accuracy, halving the prediction error to about 9\%. Learning curves are shown in the SM. 

\begin{figure}[tbhp]
\centering
\includegraphics[width=0.5\textwidth]{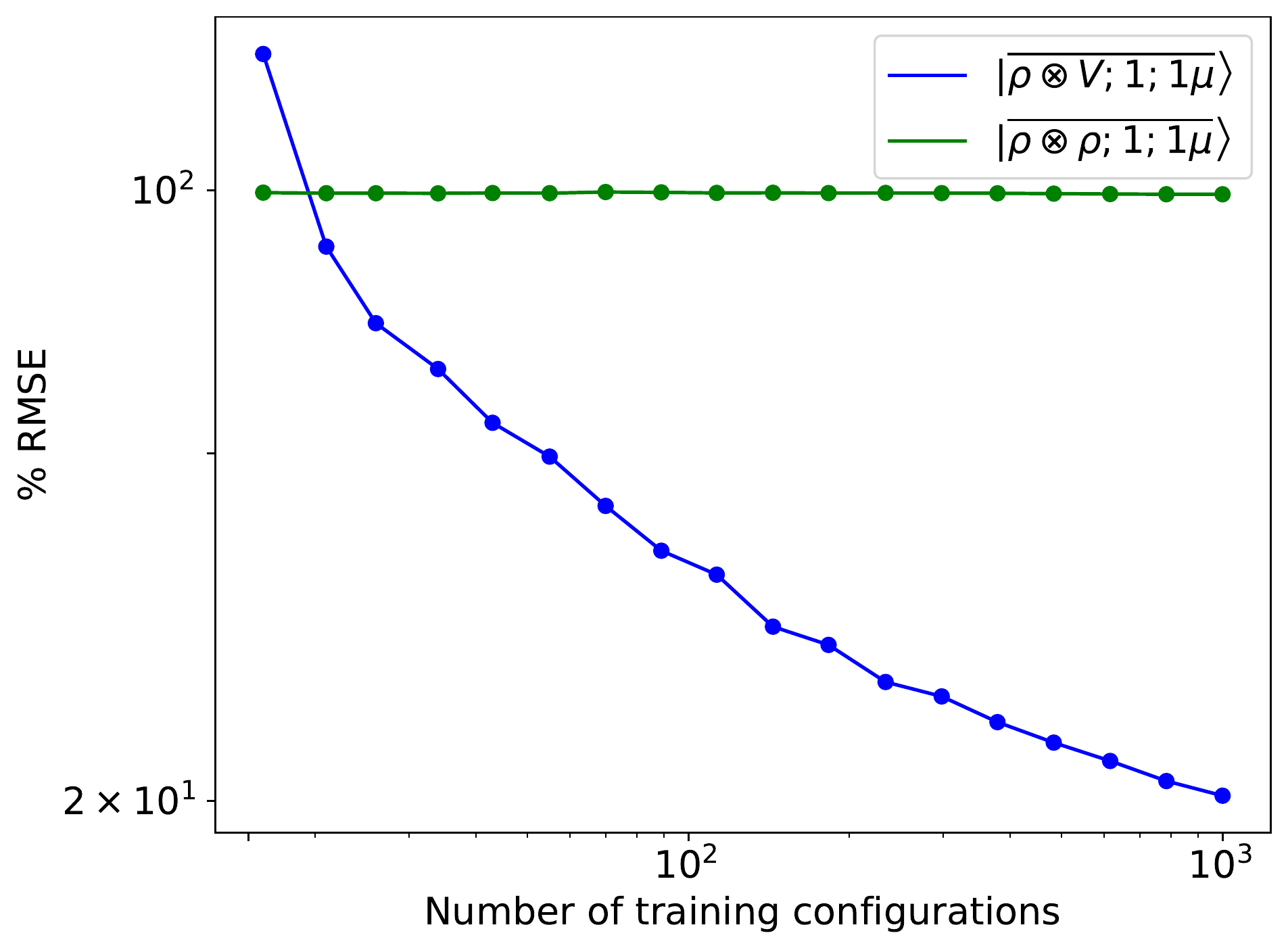}
\caption{Learning curves for the induced polarization of the water molecule due to interaction with image charges in the metal slab, computed only for separations greater than 4.5\AA{}. The error is computed as a fraction of the intrinsic variability of the test set of 215 configurations. Contrary to the local model (green), a linear $\fket{\rrhovi}$ model (blue) can learn this self-consistent polarization, with no significant reduction of the learning rate up to 1000 training configurations.}
\label{fig:polarization-lc}
\end{figure}

To further investigate what aspects of the physics of the molecule-surface interaction can be captured by the model, we perform a Mulliken population analysis on the reference DFT calculations, to extract the polarization vector of the water molecule in response to the interaction with the metal, i.e., $\bP^W = \bmu^W - \bmu^W_0$, where $\bmu^W$ and $\bmu^W_0$ are the dipole moment of the water molecule in the lithium-slab system and in vacuum respectively. 
Physically, the polarization $\bP^W$ involves the response of water's electrons to the rearrangement of the electronic charge in the surface triggered by the dipolar field, and so it involves explicitly a back-reaction. Furthermore, the polarization shows both a (usually larger)  component along the $z$-axis, and a tangential component in $xy$-plane. To account for the vectorial nature of $\bP^W$, we take advantage of the tensorial extension of Eq.~\eqref{eq:equivariant}.
To single out the long-range nature of the polarization interaction, we restrict the regression of $\bP^W$ to water configurations that are more than 4.5~{\AA} far from the surface. Our dataset contains 1215 such configurations, out of which we randomly select  1000 for training, while the remaining 215 are retained for testing.
Given that the training set contains no structures within the local descriptor cutoff, it comes as no surprise that a pure density-based tensor model  $\ket{\overline{\rho_i^{\otimes 2}; 1; 1 \mu}}$  entirely fails to learn the long-range polarization induced on the water molecule. 
Making use of the potential-based tensor model of Eq.~\eqref{eq:tensors}, in contrast, allows us to effectively learn the polarization vector $\bP^W$, showing an error that decreases to $\sim$20 \%RMSE at the maximum training set size available (Figure~\ref{fig:polarization-lc}).
This example provides a compelling demonstration of the ability of $\ket{\rrhovi}$ to build models of effects that go well-beyond permanent electrostatics. 

\begin{figure}[tbhp]
\centering
\includegraphics[width=1.0\linewidth]{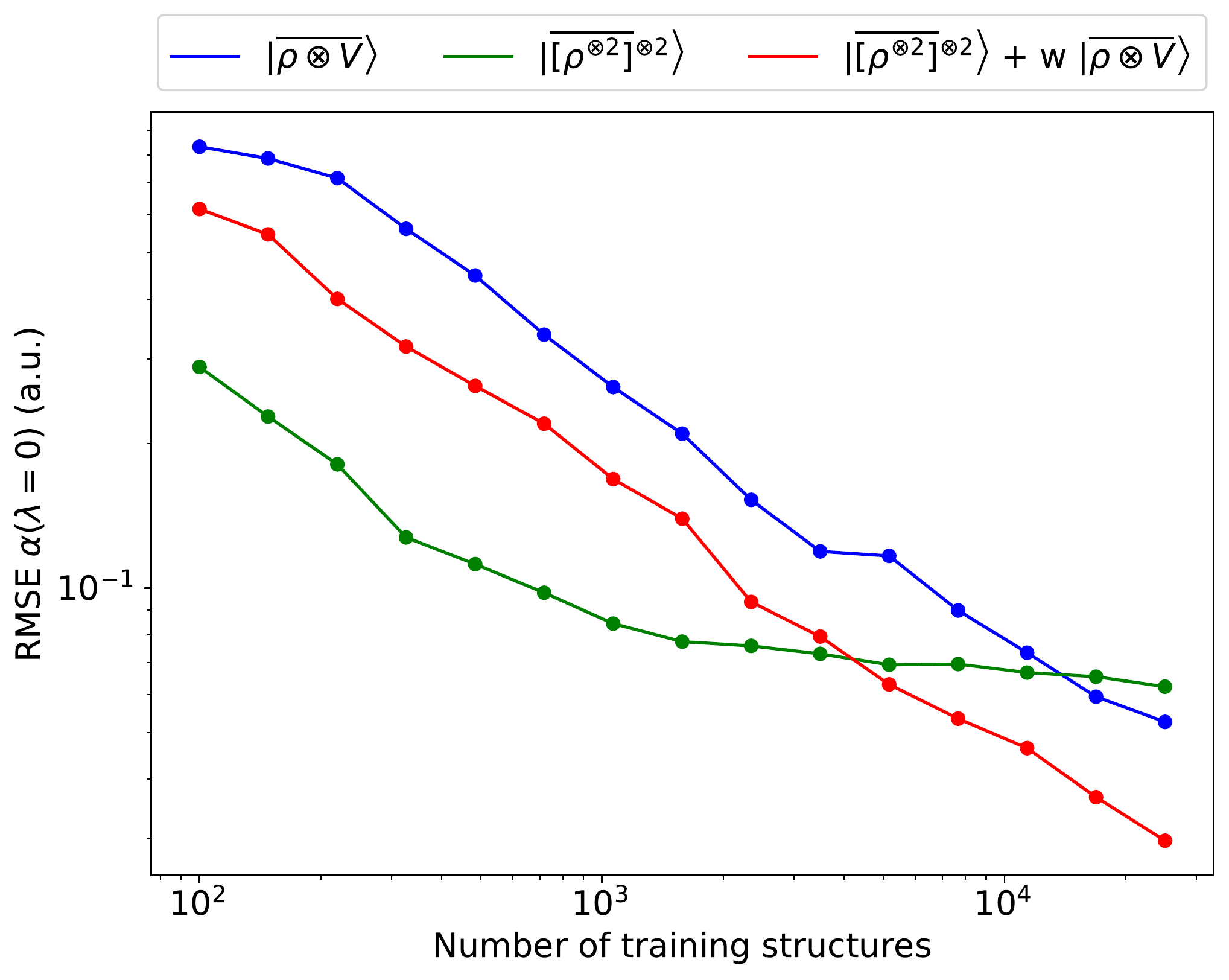}
\caption{Learning curves for the $\lambda=0$ component of the polarizability tensor of a database of polypeptide conformers.
The green curve corresponds to the non-linear kernel which is equivalent to $\fket{[\rrhoi{2}]^{\otimes 2}}$, the blue curve to a linear kernel based on $\fket{\rrhovi}$, and the red one to an optimal linear combination of the two.
\label{fig:lcurves-polypetides-alpha0} }
\end{figure}

\subsection{Response functions of oligopeptides}

As a final example, we consider the challenging task of predicting the polarizabilty of a dataset of poly-aminoacids. Dielectric response functions are strongly affected by long-range correlations, because of the cooperative nature of the underlying physical mechanism. Poor transferability of local models between structures of different sizes has been observed for molecular dipole moments~\cite{veit+20jcp}, polarizability~\cite{wilk+19pnas}, and the electronic dielectric constant of bulk water~\cite{gris+18prl}.   %
For this purpose, we use a training set composed of 27428 conformers of single aminoacids and 370 dipeptides, testing the predictions of the model on a smaller test set containing 30 dipeptides, 20 tripeptides, 16 tetrapeptides and 10 pentapeptide configurations. Reference polarizability calculations are carried out with the Gaussian16 quantum-chemistry code using the double-hybrid DFT functional PWPB95-D3 and the aug-cc-pVDZ basis set~\cite{linneathesis}.
We compute the multi-scale $\ket{\rrhovi}$ features and their local counterparts using a Gaussian width of $\sigma=$0.3~\AA{} and a spherical environment cutoff of $\rcut =$4~\AA{}. 
This data set is interesting, because it combines large structural variability with tens of thousands of distorted aminoacid configurations with longer-range interactions described by a few hundred dipeptide conformers.  
We achieve greater flexibility in the description of the local part using a square kernel model, that is equivalent to using a quadratic functional of the SOAP features, $\ket{[\rrhoi{2}]^{\otimes 2}}\equiv \ket{\rrhoi{2}}\otimes\ket{\rrhoi{2}}$. For the multi-scale LODE model, instead, we use a simple linear model, that is less susceptible to overfitting.

\begin{figure}[tbhp]
    \centering
\includegraphics[width=8.cm]{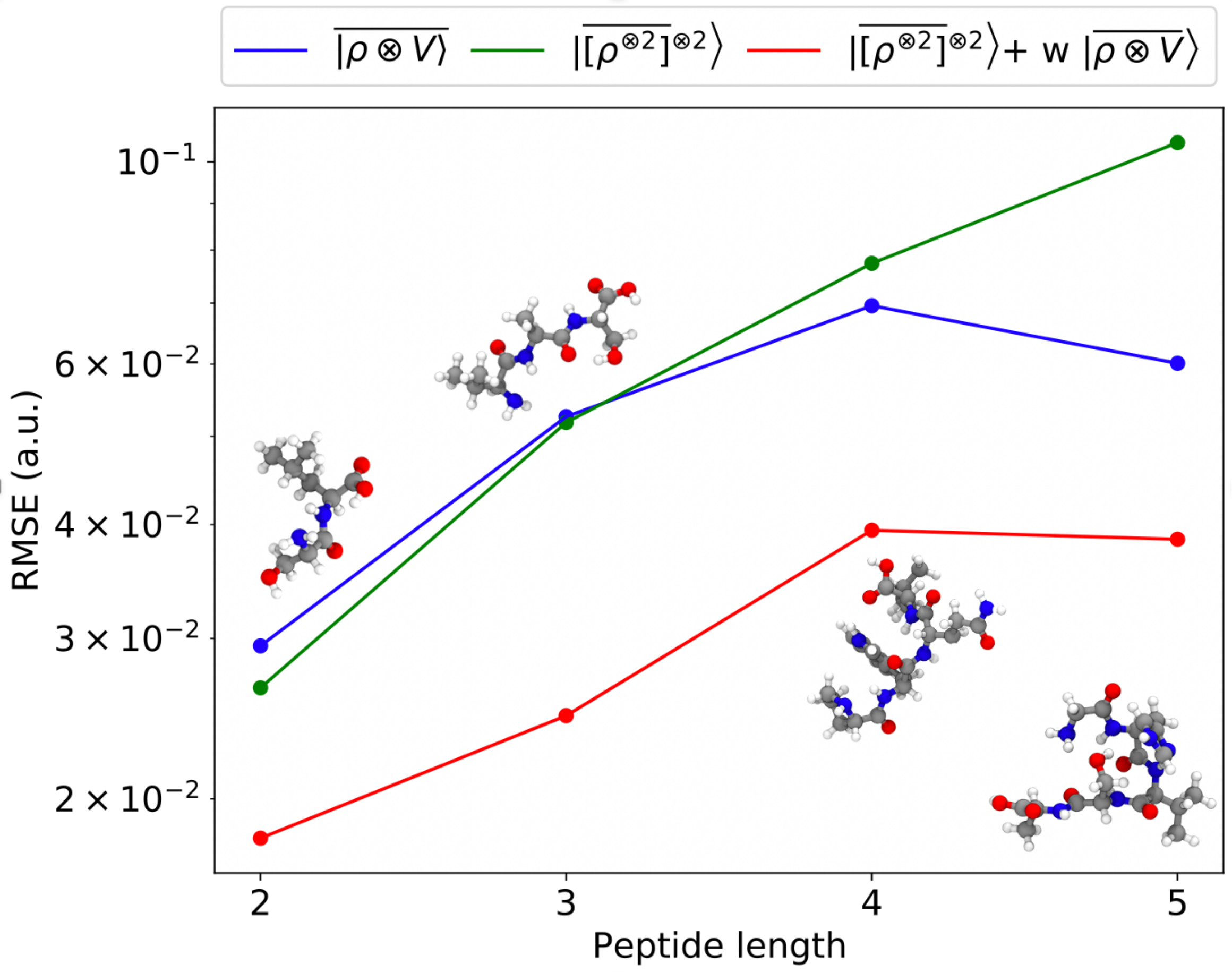}
\caption{Absolute RMSE in learning the $\lambda=0$ spherical tensor of polarizability of polypeptides as a function of the peptide length. The model was trained on 27428 single-amino acids and 370 dipeptides. The error was computed on 30 dipeptides, 20 tripeptides, 16 tetrapeptides and 10 pentapeptides respectively.}
    \label{fig:abs-rmse-polypetide-length-alpha0}
\end{figure}

The learning curves for the trace ($\lambda=0$) of the polarizability tensor, shown in Fig.~\ref{fig:lcurves-polypetides-alpha0}, are very revealing. 
The $\ket{[\rrhoi{2}]^{\otimes 2}}$ model, which disregards any non-local behavior beyond the atomic environment, is initially very efficient, but saturates to an error of 0.06a.u.. In contrast, equipped with non-local information, the  $\ket{\rrhovi}$ representation reduces the error of prediction to 0.05a.u., but is initially much less effective. 
This is not due to the lack of higher-order local density correlations: a linear $\ket{\rrhoi{2}}$ model performs well, despite showing saturation due to its local nature (see discussion in the SM).
Rather, it is indicative of the importance of short-range effects in this diverse dataset. Inspired by Refs.~\cite{bart+17sa,will+18pccp,paru+18ncomm} we build a tunable kernel model based on a weighted sum of the local and the LODE kernels.
We optimize the weight by cross-validation at the largest train size, obtaining a reduction of 50\%{} of the test error, down to 0.028a.u.
An analysis of the test error which separates the contributions from oligopeptides of different length, shown in Fig.~\ref{fig:abs-rmse-polypetide-length-alpha0}, is consistent with this interpretation of the learning curves. 
All models show an error that increases with the size of the molecule, because there are interactions that are just not described at the smaller train set size. 
However, the purely local model shows by far the worst extrapolative performance, while multi-scale models -- in particular the one combining a non-linear local kernel and LODE features -- show both a smaller overall error, and a saturation of the error for tetra and penta-peptides. 
This example illustrates the different approaches to achieve a multi-scale description of atomic-scale systems: the $\ket{\rrhovi}$ features offer simplicity and physical interpretability, while a multi-kernel model makes it possible to optimize in a data-driven manner the balance between local and long-ranged correlations. 

\section{Conclusions}

The lack of a description of long-range physical effects is one of the main limitations of otherwise greatly successful machine-learning schemes which are more and more often applied to model atomic-scale phenomena. 
We show how it is possible to construct a family of multi-scale equivariant features, that combine the properties of well-established local ML schemes with the long-distance equivariant features that have been recently proposed by some of the Authors. 
This multi-scale framework shows enticing formal correspondences with physically-meaningful interaction terms, such as multipole electrostatics. Still, the data driven nature of the construction allows the description of long-range interactions that do not fit this specific physical model.
We show examples of how a multi-scale LODE model can accurately predict interactions between different kinds of molecular dimers, that include charged, polar and apolar compounds.
Results are also very promising when it comes to modelling systems that clearly go beyond permanent electrostatics, such as a water molecule interacting with a metallic slab, and the dielectric response of oligopeptides. 
The combination of a physics-inspired formulation and data-driven flexibility that underlies this multi-scale LODE framework addresses one of the outstanding issues in atomistic machine-learning, and paves the way towards an even more pervasive use of statistical methods to support the computational investigation of molecules and condensed-phase systems. 

\section*{Acknowledgments}

The Authors would like to thank David Wilkins and Linnea Folkmann for sharing training data for the polarizability of polyaminoacids, and Mariana Rossi for help computing DFT references for the water-surface binding curves.
M.C and A.G. were supported by the European Research Council under the European Union's Horizon 2020 research and innovation programme (grant agreement no. 677013-HBMAP), and by the NCCR MARVEL, funded by the Swiss National Science Foundation. A.G. acknowledges funding by the MPG-EPFL centre for Molecular Nanoscience and Technology. JN was supported by a MARVEL INSPIRE Potentials Master's Fellowship. 
We thank CSCS for providing CPU time under project id s843.

\end{document}